\documentclass[a4paper,12pt]{article}

\usepackage{a4wide}
\usepackage{graphicx,subfigure,hhline}
\usepackage[colorlinks=true,linkcolor=blue,citecolor=red]{hyperref}
\usepackage{xcolor}
\usepackage{amssymb,amsthm,empheq,bbold}
\usepackage[inline]{enumitem}
\usepackage[ruled,vlined,linesnumbered]{algorithm2e}
\usepackage{caption}
\theoremstyle{plain} 
\newtheorem{theorem}{Theorem} 

\newtheorem{remark}{Remark}
\newtheorem{definition}{Definition}

\allowdisplaybreaks

\begin{document}
	
	\title{A nonconservative kinetic model under the action of an external force field for modeling the medical treatment of autoimmune response}

	\author{Marco Menale{$^{1*}$, Romina Travaglini{$^{2,3}$}}\\[1em]
		$^1${\footnotesize Department of Mathematics and Applications ``R. Caccioppoli",}\\ {\footnotesize  University of Naples ``Federico II",}\\ {\footnotesize  Via Cintia, Monte S. Angelo I-80126 Napoli, Italy} \\{\footnotesize marco.menale@unina.it (*corresponding author)}
		\\[0.5em]
		$^2${\footnotesize INDAM -- National Institute for Advanced Mathematics  ``Francesco Severi''}
		\\{\footnotesize  Piazzale Aldo Moro 5,  00185, Roma, Italy}
		\\{\footnotesize travaglini@altamatematica.it  }
			\\[0.5em]
		$^3$
		 {\footnotesize Department of Mathematical, Physical and Computer Science, University of Parma}\\{\footnotesize Parco Area delle Scienze 53/A,
			43124, Parma, Italy}}
	
	\date{}
	\maketitle

	\begin{abstract}
		
		In this paper, we develop a nonconservative kinetic framework to be applied to the study of immune system dysregulation. From the modeling viewpoint, the model regards a system composed of stochastically interacting agents, under the action of an eternal force field. According to the application perspectives of this paper, the external force field has a specific analytical shape. In this case, some analytical results are proved, i.e. existence, uniqueness, positivity, and boundedness of solution of the related Cauchy problem, at least locally in time. Then, the model is refined to be implemented for the study of treatment strategies in case of autoimmune response. Specifically, we distinguish the autonomous case from the nonautonomous one, representing the absence or delivery of drugs, respectively. The former allows us to gain some stability results. Whereas, the latter is qualitatively studied. Numerical simulations are provided for both schemes.
	\end{abstract}
	
	\smallskip
	
	
	\smallskip
	
	
	\maketitle

	\section{Introduction}

	Kinetic theory for interacting particles has been widely studied and used in the last decades (see \cite{bellomo2009complexity}, and references therein, for a complete description). From the kinetic viewpoint, an interacting system is seen as composed of \textit{particles}, also called \textit{agents}. These agents stochastically interact during the evolution of the system; the interactions are binary and modeled by some parameters. A multiscale approach is used to model the dynamics. At a \textit{microscopic level}, the state of particles is described by \textit{mechanical} and \textit{nonmechanical} real variables. In particular, the nonmechanical one is called \textit{activity}. In the current paper, only the latter characterizes the microscopic state of the system. The overall system is divided into functional subsystems (see \cite{bellomo2009complexity, bianca2020mathematical} for details), such that particles belonging to the same functional subsystem share the same strategy. The meaning of this ``strategy" depends on the particular application taken into account. The introduction of \textit{distribution functions}, related to the functional subsystems, furnish the \textit{mesoscopic description} of the system. Therefore, at this level, we are not interested in the specific evolution of single particles, but in the evolution of each functional subsystem. Roughly speaking, it is a ``statistical description" of the system, since the evolution of functional subsystems is modeled in terms of distribution functions. Depending on the particular shape of the microscopic variable, i.e. the activity, the mesoscopic level is described by integro-differential equations, partial-differential equations, or ordinary differential equations. In this paper, the activity variable attains its values in a continuous real subset, therefore the evolution of the system is described by a system of nonlinear integro-differential equations. Finally, the introduction of suitable moments, obtained by integrating distribution functions with respect to the microscopic variable, the \textit{macroscopic equations} of the system are derived. These equations furnish the overall state of the system, with the related dynamics.
	
	It is worth stressing that kinetic theory modeling approach has been widely used in the last decades, due to its versatility. For instance, there are applications in economy \cite{bertotti2011microscopic, bertotti2014micro}, vehicular traffic \cite{fermo2015fully}, psychology \cite{bellomo2008complexity}, biology \cite{conte2022multi}, and opinion dynamics \cite{lachowicz2019diffusive}. Moreover, the recent COVID-19 pandemic event has increased interest in the modeling of epidemiology by kinetic theory models. Among the others, in terms of epidemiological dynamics \cite{dimarco2021kinetic, della2023sir}, behavioral epidemiology \cite{della2022intransigent}, and socio-economical impact and consequences \cite{dimarco2020wealth}. In medical frameworks, kinetic models have been used to study dynamics of tumor growth describing the interplay between cancer cells and the immune system \cite{bellomo1994dynamics}, and then extended in different ways, also including spatial motion \cite{bellomo2004class}. Moreover, according to the specific aim of this paper, these tools have been applied also to the study of autoimmune diseases \cite{delitala2013mathematical, della2022mathematical, kolev2019mathematical, kolev2018mathematical,ramos2019kinetic}, with the further extension to time-spatial case \cite{immuneProc}. 
	
	From an analytical viewpoint, conservative kinetic frameworks ensure some analytical properties of solution, i.e. existence and uniqueness globally in time, with positivity and boundedness. These latter properties are important in order to obtain a physically realistic solution. Therefore, the introduction of nonconservative terms may cause the loss of, at least, one of the previous properties. Nevertheless, for some applications, nonconservative interactions cannot be neglected if a more realistic and authentic description is required. Then, proliferative/destructive events or the action of the external environment is crucial for the evolution of some stochastically interacting systems. Therefore, this paper regards the development and analysis of a kinetic framework under the action of an external force field. {In particular, the introduction of an external force field is related to what has been recently done, for the discrete case, in the paper \cite{carbonaro2023nonconservative}, motivated by an ecological application, and in the  paper \cite{menale2023kinetic}, motivated by an epidemiological application}. This choice ensures, at least locally in time, the existence and uniqueness of a bounded and positive solution. Moreover, further assumptions can provide the extension of the solution globally in time.

	The kinetic setting presented above turns out to be suitably applicable to the autoimmune response in the presence of medical treatment. Dysfunctions of the human immune system are undoubtedly one of the most investigated medical research fields. The protection mechanism played by the immune system against external pathogen agents, as well as its role in repairing and regeneration of damaged tissues, is as important as complicated to understand. Unfortunately, a wide range of possible dysregulation in immune dynamics may occur, such as autoimmune disorders or immunodeficiency conditions. In the case of autoimmune syndromes, they arise when immune cells manage to react against human tissues themselves, without being suppressed by specific regulation mechanisms. This may lead to different severe and debilitating surges. Causes of such syndromes are not clear yet, they may be genetic \cite{somers2006autoimmune}, environmental \cite{hess2002environmental}, or even viral \cite{marchewka2023rare}. On the other hand, the extreme variance of possible dynamics leading to an autoimmune attack makes a still open challenge for both description and treatment of these pathological conditions (see \cite{wang2015human}, and references therein).

	The main process underlying autoimmune disorders is the fact that naive lymphocytes (as T-cells or B-cells), able to react against antigens presented by cells of the human body, are not suppressed during thymic maturation (negative selection), as would happen in healthy patients. Consequently, they manage to reach peripheral lymphoid tissues and become activated once they encounter self-cells presenting their cognate antigen. This leads to an inflammation of target tissue and cytokines production, that stimulate activation and proliferation of self-reactive lymphocytes, starting the autoimmune cascade. On the other hand, migration of self-reactive lymphocytes in peripheral tissues is possible also in healthy patients, \cite{danke2004autoreactive}. Therefore, populations of ISCs, such as regulatory T lymphocytes (Tregs) and natural killer cells, have the task of inhibiting or inducing apoptosis of self-antigen presenting cells and self-reactive lymphocytes. 
	
	To build our model, we refer to the framework for autoimmunity proposed in \cite{ramos2019kinetic}, where a kinetic model is presented to describe interaction dynamics among self-antigen presenting cells (SAPCs), self-reactive T-cells (SRTCs), and immunosuppressive cells (ISCs). This model has been extended including also the impact of external factors as a constant input of SAPCs due to environmental agents or unhealthy habits  \cite{della2022mathematical}, or a drug therapy consisting of an intake of Interleukin-2 cytokines \cite{costa2021optimal}. 
	
	In the cited cases, though, a study of the analytical properties of the models proposed is performed only for the specific cases treated. For this reason, we found it appropriate to provide the first analytical results for the general case. Moreover, in the references above, interactions between two populations of cells may be conservative, proliferative, or destructive. Specifically, in the conservative case, cells change their internal activity, still remaining in the same functional subsystem. In the present case, we consider both naive and activated immune cells and a conservative (for the total population of immune cells) process of activation. More precisely, 
	we focus on the self-reactive B-cells (SRBs) population, that are known to play a relevant role in autoimmune disease development \cite{mamula1994b,harris2000reciprocal}, and we split the SRBCs population into two subsystems: naive B-cells (NBCs), being B-cells able to react against self-antigens, but have not been activated yet, and memory self-reactive B-cells (MSBCs), that come from the activation of NBCs when encounters with SAPCs occur, \cite{dorner2009b}.
	For the population of ISCs, we only consider Tregs, which are a subset of CD4$^+$ T lymphocytes whose importance against autoimmune response is widely known, \cite{akkaya2019regulatory,liberal2015regulatory}. Whereas,
	the population of SAPCs is kept general, since it may involve different types of cells, depending on the peculiar autoimmune disease. 
	
	In addition, we point out that a relevant feature of some autoimmune diseases is their chronic course, i.e. they exhibit relapsing-remitting dynamics, during which periods of appearing and disappearing of symptoms alternate  \cite{rosenblum2015mechanisms}. This recurrent behavior has been investigated mathematically in \cite{zhang2014modeling} and \cite{zhang2016hopf} through a Hopf bifurcation of the solution for macroscopic system, and then derived from the kinetic level in \cite{della2022mathematical}. 
	
	According to the theoretical framework proposed, we model here a treatment strategy in the form of an external force field. There are many different strategies adopted in medicine to regulate the anomalous immune response in the human body, and they vary with respect to the particular autoimmune disease they are applied to (see \cite{chandrashekara2012treatment,serra2015nanoparticle,shin2018treatment}, and references therein). In this case, we focus on Rituximab and low dose of interleukin L-2  (IL-2). Rituximab is a specific antibody able to attack CD20 (a glycoprotein present on the surface of B cells), thus it can cause apoptosis of both NBCs \cite{kamburova2012vitro,szodoray2004apoptotic} and MSBCs \cite{gottenberg2005tolerance}. Whereas, IL-2 plays a crucial role in autoimmune tolerance. Indeed, infusion of low doses IL-2 may stimulate expansion of Tregs \cite{klatzmann2015promise}, and enhance their suppressive function \cite{furtado2002interleukin}. Many medical studies have been carried out on the treatment of patients and results for both Rituximab \cite{gottenberg2005tolerance} and low-dose IL-2 cases \cite{koreth2011interleukin}. We aim to adapt our model to reproduce the therapy delivery by means of an external force field and compare qualitatively numerical results with those expected in the medical framework. 	
	
	The contents of this paper are organized as follows. Section \ref{secmodel} provides the general nonconservative kinetic model under the action of a particular shape external force field. Some preliminary analytical results are gained in Section \ref{secanalyt}. In Section \ref{secappl}, the application of the kinetic framework in the context of autoimmune response is presented and discussed. After specializing the kinetic framework for this problem in Subsection \ref{subsectkin}, some stability considerations are provided in Subsection \ref{Subsecmacro}, with respect to the macroscopic system. Furthermore, numerical simulations are performed in Subsection \ref{Subsecnum}. Finally, conclusions and research perspectives are discussed in Section \ref{seconcl}.

	\section{The kinetic model}\label{secmodel}
	
	Let us consider an interacting system composed of \textit{agents}, also called \textit{particles}, \textit{homogeneous} with respect to \textit{mechanical variables}, i.e. \textit{space} and \textit{velocity}. The interaction among particles is binary and stochastic, and it is ruled by some parameters, whose specific meaning depends on the particular application, as widely shown in the following.
	
	The system is divided into \textit{functional subsystems} \cite{bellomo2009complexity}, which may be seen as particular subsets of particles. For instance, according to the main aims of this paper, in a medical setting, functional subsystems can correspond to populations of cells sharing the same biological role.
	
	The \textit{microscopic state} of the system is defined by a real scalar variable $u \in \mathbb{R}$, called \textit{activity}. In particular, 
	$$u \in D_u\subseteq \mathbb{R},$$ 
	where $D_u$ is a bounded real subset.
	
	As well as for functional subsystems, the meaning of the activity $u$ depends on the particular application. For the application here considered, the activity of cells can represent their level of activation and efficiency in performing their task.

	Hereafter, we suppose that the overall system is divided into $n \in \mathbb{N}$ functional subsystems. Then, the \textit{distribution function} over the $i$th functional subsystem, for $i \in \{1, 2, \dots, n\}$, reads 
	$$f_i(t,u):[0,\, T]\times D_u \rightarrow \mathbb{R}^+, \qquad T>0.$$
	Moreover, the \textit{vector distribution function} is
	$$\mathbf{f}(t)=\left(f_1(t),\, f_2(t),\, \dots, \, f_n(t)\right).$$
	Two classes of parameters describe the stochastic microscopic dynamics between pairs of particles. Specifically:
	
	\begin{itemize}
		\item The \textit{transition probability}
		$$\mathcal{A}_{hk}^i(u,\,u_*,\,u^*):D_u\times D_u \times D_u \rightarrow \mathbb{R}^+,$$
		for $i,h,k \in \{1, 2, \dots, n\}$, that is the probability that a particle of the $h$th functional subsystem in the microscopic state $u_*$ falls into the state $u$ of the $i$th functional subsystem after the interaction with a particle into of the $k$th functional subsystem in the microscopic state $u^*$. As we require it is a probability, the following assumption holds:
		\begin{equation}\label{assp}
			\sum_{i=1}^n\int_{D_u}\mathcal{A}_{hk}^i(u,\,u_*,u^*)\, du=1, \qquad \forall h,k\in\{1, 2, \dots, n\},\, \forall u_*,u^*\in D_u.
		\end{equation}
		\item The \textit{interaction rate}
		$$\eta_{hk}(u_*,\,u^*):D_u\times D_u \rightarrow \mathbb{R}^+,$$
		for $h,k \in \{1, 2, \dots, n\}$, that is the number of encounters per unit of time between a particle in the microscopic state $u_*$ of the $h$th functional subsystem and a particle in the microscopic state $u^*$ of the $k$th functional subsystem. 
	\end{itemize}
	
	\noindent Bearing all above in mind, the \textit{kinetic equation} that describes the evolution of the $i$th functional subsystem at a \textit{mesoscopic level}, for $i\in \{1, 2, \dots, n\}$, is
	\begin{equation}\label{kineq}
		\frac{\partial}{\partial\,t}f_i(t,u)=G_i[\mathbf{f}](t,u)-L_i[\mathbf{f}](t,u),
	\end{equation}
	where
	\begin{equation}\label{kineqGi}G_i[\mathbf{f}](t,u):=\sum_{h,k=1}^{n}\int_{D_u\times D_u}\eta_{hk}(u_*,u^*)\mathcal{A}_{hk}^i(u,u_*,u^*)f_h(t,u_*)f_k(t,u^*)\,du_*du^*,\end{equation}
	for $i\in \{1, 2, \dots, n\}$, is the \textit{gain-term operator}
	that gives the number of particles that enter into the $i$th functional subsystem and acquire the state $u$, and
	\begin{equation}\label{kineqLi}L_i[\mathbf{f}](t,u):=f_i(t,,u)\sum_{k=1}^n\int_{D_u}\eta_{ik}(u,u^*)f_k(t,u^*)\, du^*,\end{equation}
	for $i \in \{1, 2, \dots, n\}$, is the \textit{loss-term operator} that gives the number of particles in the microscopic state $u$ that leave the $i$th functional subsystem.

	Roughly speaking, the right-hand side of the kinetic equation \eqref{kineq} is a net-flux related to the particles of the $i$th functional subsystem, for $i\in\{1, 2, \dots, n\}$, in the microscopic state $u$. Due to the previous assumption \eqref{assp}, the model \eqref{kineq} is conservative. Indeed, By integrating on $D_u$ and summing on $i \in\{1,2, \dots, n\}$, one has
	\begin{equation}\label{eqcon}
		\frac{d}{d\,t}\left(\sum_{i=1}^n\int_{D_u}f_i(t,u)\, du\right)=0,
	\end{equation}
	that is the total amount of particles in the overall system does not vary during the evolution, because neither proliferative/destructive events nor external actions occur. 
	
	The \textit{$p$th-order moment of the system}, for $p \in \mathbb{N}$, defines
	$$\mathbb{E}_p[\mathbf{f}](t):=\sum_{i=1}^n\int_{D_u}u^p\,f_i(t,u)\, du,$$
	where $\displaystyle \int_{D_u}u^p\,f_i(t,u)\, du$ is the \textit{local $p$th-order moment} of the system. The $p$th-order moment, for $p \in \mathbb{N}$, describes the \textit{macroscopic state}. For instance, from a physical viewpoint, the $0$th-order,  $1$st-order, and $2$nd-order represent  \textit{density}, \textit{linear momentum} and \textit{global activation energy}, respectively. Then, the property \eqref{eqcon} is equivalent to the conservation of the \textit{global density} of the system, that we indicate by $\rho(t)$. Moreover, if the (local) density of the $i$th functional subsystem, for $i \in \{1, 2, \dots, n\}$, is defined as
	\begin{equation}\label{rhoi}
		\rho_i(t):=\int_{D_u}f_i(t,u)\,du,\end{equation}
	then the global density rewrites
	$$\rho(t)=\sum_{i=1}^{n}\rho_i(t).$$
	Finally, the property \eqref{eqcon} ensures the conservation of the global density $\rho(t)${, i.e.
	$$\frac{d\rho(t)}{dt}=0.$$
	Nevertheless, this property does not extend to each functional subsystem, i.e. the local density $\rho_i(t)$, for $i \in \{1, 2, \dots, n\}$, can change during the evolution of the system.}

	The conservative kinetic framework \eqref{kineq} does not consider any external action over the system, and this may be realistic for \textit{closed} systems. Nevertheless, in some applications, the action of the environment over the system cannot be neglected for a more realistic description, according to the aims of the current paper. This justifies the introduction of an \textit{external force field} into the kinetic framework \eqref{kineq}. Generally, this external force field depends on time, activity variable, and distribution function $f_i(t,u)$ of the $i$th functional subsystem, for $i \in \{1, 2, \dots, n\}$. Therefore, for $i \in\{1, 2, \dots, n\}$,
	$$F_i[\mathbf{f}](t,u):\left(\mathbb{R}^+\right)^n \times [0,\, T] \times D_u \rightarrow \mathbb{R}$$
	is the $i$th component of the external force field. The \textit{vector external force field} writes
	$$\mathbf{F}[\mathbf{f}](t,u):=\left(F_1[\mathbf{f}](t,u), F_2[\mathbf{f}](t,u), \dots, F_n[\mathbf{f}](t,u)\right).$$ 
	Bearing the conservative framework \eqref{kineq} in mind, the kinetic equation that describes the evolution of the $i$th functional subsystem, under the action of the external force field $\mathbf{F}[\mathbf{f}](t,u)$, writes
	\begin{equation}\label{forckin}
		\frac{\partial}{\partial\,t}f_i(t,u)=G_i[\mathbf{f}](t,u)-L_i[\mathbf{f}](t,u)+F_i[\mathbf{f}](t), \qquad i \in\{1, 2,\dots, n\}.
	\end{equation}
	In general, the external force field $\mathbf{F}[\mathbf{f}](t,u)$ has a specific analytical shape according to the particular application taken into account. For our current aim, the $i$th component of the external force field, for $i\in \{1, 2, \dots, n\}$, has the following shape 
	\begin{equation}\label{forshap}
		F_i[\mathbf{f}](t,u)= f_i(t,u)\left(\sum_{h=1, h \neq i}^{n}\int_{D_u}\Gamma_{i h}(t)\, f_h(t,u) \,d u+ \Theta_i (t)  \right)+\,\alpha_i,
	\end{equation}
	where $\Theta_i(t)$, for $i \in \{1, 2, \dots, n\}$ and $\Gamma_{ih}(t)$, for $i,h \in \{1, 2, \dots, n\}$, are real-valued functions, and $\alpha_i \geq 0$, for all $i \in \{1, 2, \dots, n\}$. Now, according to the definition of local density $\rho_h(t),\,h \in \{1, 2, \dots, n\}$, given in \eqref{rhoi},
	equation \eqref{forshap} can be rewritten as
	\begin{equation}\label{forshap2}
		F_i[\mathbf{f}](t,u)= f_i(t,u)\left(\sum_{h=1, h \neq i}^{n}\Gamma_{i h}(t)\, \rho_h(t)+ \Theta_i (t)  \right)+\alpha_i,
	\end{equation}
	or, equivalently,
	\begin{equation}\label{forshap3}
		F_i[\mathbf{f}](t,u)= f_i(t,u)\sum_{h=1, h \neq i}^{n}\Gamma_{i h}(t)\, \rho_h(t)+ f_i(t,u)\Theta_i (t)+\alpha_i.
	\end{equation}
	The $i$th component of the external force field \eqref{forshap3}, for $i \in \{1, 2, \dots, n\}$, can be seen as composed of three terms: $f_i(t,u)\sum_{h=1, h \neq i}^{n}\Gamma_{i h}(t)\, \rho_h(t)$, $ f_i(t,u)\Theta_i (t)$ and $\alpha_i$. The first term, $f_i(t,u)\sum_{h=1, h \neq i}^{n}\Gamma_{i h}(t)\, \rho_h(t)$, expresses the dependence of this action with respect to the local density of the other functional subsystems, weighted by the function $\Gamma_{ih}(t)$. The second term, $ f_i(t,u)\Theta_i (t)$, refers to the \textit{direct} action of the external force field on the $i$th functional subsystem itself. The presence of the term $f_i(t,u)$ reveals the dependence of $F_i[\mathbf{f}](t,u)$ on the microscopic state $u$. Finally, the third term, $\alpha_i$, can be seen as a constant external resource for $i$th functional subsystem. In particular, if $\alpha_i=0$ and there are no particles in the $i$th functional subsystem with microscopic state $u$ at time $t>0$, i.e. $f_i(t,u)=0$, then $F_i[\mathbf{f}](t,u)=0$, that is there are no consequences due to the external action. Roughly speaking, by using the shape \eqref{forshap}, the external force field acts \textit{directly} on each functional subsystem, this action depends also on the rest of the system, and it can also provide a constant resource.
	
	Bearing all above in mind, the \textit{kinetic equation with external force field} modeling the dynamics of the $i$th functional subsystem, for $i \in \{1, 2, \dots, n\}$, reads
	\begin{equation}\label{eqfkin}\begin{split}
			\frac{\partial}{\partial\,t}f_i(t,u)&=\sum_{h,k=1}^{n}\int_{D_u\times D_u}\eta_{hk}(u_*,u^*)\mathcal{A}_{hk}^i(u,u_*,u^*)f_h(t,u_*)f_k(t,u^*)\,du_*du^*\\&
			-f_i(t,,u)\sum_{k=1}^n\int_{D_u}\eta_{ik}(u,u^*)f_k(t,u^*)\, du^*\\&
			+f_i(t,u)\left(\sum_{h=1, h \neq i}^{n}\int_{D_u}\Gamma_{i h}(t)\, f_h(t,u^*) \,d u^*+ \Theta_i (t)  \right)+\alpha_i.
	\end{split}\end{equation}
	In general, this kinetic framework \eqref{eqfkin} does not ensure the conservation of the global density $\rho(t)$, i.e.
	$$\frac{d\rho}{dt}(t)\neq 0, \qquad t>0,$$
	since the external force field may determine \textit{proliferative} or \textit{destructive} events.
	
	If a suitable \textit{initial data} is assigned, i.e.
	$$\mathbf{f}^0(u)=\left(f^0_1(u), f^0_2(u), \dots, f^0_n(u)\right),$$
	the \textit{initial value problem} or \textit{Cauchy problem} related to the framework \eqref{eqfkin} defines
	\begin{equation}\label{ivp}
		\begin{cases}
			\eqref{eqfkin} &{ (t,u)\in [0,\,T]\times D_u}\\\\
			\mathbf{f}(0,u)=\mathbf{f}^0(u) &{ u \in D_u}.
		\end{cases}
	\end{equation}
	{If no external force field acts, then the above initial value problem \eqref{ivp} admits a unique and positive solution, globally in time, once a positive initial data $\mathbf{f}^0$ has been assigned. Otherwise, the presence of an external force field $\mathbf{F}[\mathbf{f}](t,u)$ does not still ensure, in general, the latter analytical properties. Indeed, neither positivity nor boundedness is guaranteed. For instance, a negative external force field can lead to negative solutions, even with a positive initial data. On the other hand, a constant and positive external force field can cause blow-up phenomena. Nevertheless, for the specific choice \eqref{forshap}, provided in this paper, some further assumptions, not too restrictive, allow to obtain a unique and positive solution, at least locally in time, as shown in next Section.}

	\section{Analytical results}\label{secanalyt} 
	This Section aims at presenting some first analytical results towards the initial value problem \eqref{ivp}, related to the kinetic framework \eqref{eqfkin}. {We adopt an approach similar to that developed in \cite{bellomo2010complexity}. Hereafter, for any \textit{function space} $X$, let $X_+$ denote the \textit{positive cone} of the positive functions in $X$. In this paper, 
		$$X=\left(L^1(D_u)\right)^n,$$
		where the space $ \left(L^1(D_u)\right)^n$ is endowed with the norm
		$$\left\|\mathbf{f}(t)\right\|_1=\sum_{i=1}^n\int_{D_u}|f_i(t,u)|\,du,$$
		for $\mathbf{f}(t,u)\in\left(L^1(D_u)\right)^n$. It is worth pointing out that, in what follows, $\|\,\|_1$ refers to the above norm.} Some suitable assumptions are mandatory for the aim of this paper (see \cite{arlotti1996qualitative} and \cite{arlotti1996solution} for details):
	{
	\begin{description}
		\item[H1] \begin{equation}\label{assp1}
			\mathbf{f}^0(u)\in L_+^1(D_u).
		\end{equation}
		\item[H2] For some $T>0$:
		\begin{align}
			&\Gamma_{ih}(t)\in C([0,\, T]), \label{asspcon1}\\
			&\Theta_i(t)\in C([0,\, T])\label{asspcon2}.
		\end{align}
		\item[H3] There exists $\eta>0$ such that
		\begin{equation}\label{assp4}
			\eta_{hk}(u_*,u^*)\leq \eta, \qquad \forall h,k \in \{1, 2, \dots, n\}, \, \forall u_*,u^* \in D_u.
		\end{equation}
	\end{description}}
	 \noindent {Therefore,} for $i \in \{1, 2, \dots, n\}$, the operator
	
	\begin{equation}\label{opersigma}
	\Sigma_i[\mathbf{f}](t,u)=G_i[\mathbf{f}](t,u)+\alpha_i
	\end{equation}
	maps $X_+$ into itself due to the assumptions \eqref{assp} and \eqref{assp4}, and the fact that $\alpha_i \geq 0$, for all $i \in \{1, 2,\ \dots, n\}$. Bearing the operator $L_i[\mathbf{f}](t,u)$ in mind, let us introduce the operator $\Omega_i[\mathbf{f}](t,u)$, for $i \in\{1, 2, \dots, n\}$, defined as
	\begin{equation}\label{operomega}
	\Omega_i[\mathbf{f}](t,u):=\left(\sum_{k=1}^n\int_{D_u}\eta_{ik}(u,u^*)f_k(t,u^*)\, du^*\right) +\left(\sum_{h=1, h \neq i}^{n}\int_{D_u}\Gamma_{i h}(t)\, f_h(t,u^*) \,d u^*+ \Theta_i (t)  \right).
	\end{equation}
	{Now, the further assumptions \eqref{asspcon1} and \eqref{asspcon2}}  ensure that the operator $\Omega_i[\mathbf{f}](t,u)$ maps $X_+$ into $X$, for $i \in \{1, 2, \dots, n\}$.
	
	Bearing all the above in mind, we can define a solution for the initial value problem \eqref{ivp} on the Banach space $X$ (for analytical details see \cite{deimling2006ordinary}, and references therein).
	\begin{definition}\label{def}
		The function $\mathbf{f}$ is a \textit{solution} of the initial value problem \eqref{ivp} if 
		$$\mathbf{f}(\cdot,u): [0,\, T] \rightarrow X_+$$
		is a continuously differential function in $[0,\, T]$, and satisfies the differential equation on the Banach space $X$
		$$\frac{d\mathbf{f}}{dt}(t)=\mathbf{C}[\mathbf{f}](t), \qquad 0\leq t <T,$$
		where the operator $\mathbf{C}[\mathbf{f}](t)$ is defined as follows
		$$(\mathbf{C}[\mathbf{f}])_i(t):=\Sigma_i[\mathbf{f}](t,u)-f_i(t)\Omega_i[\mathbf{f}](t,u), \qquad i \in \{1, 2, \dots, n\},$$
		and the initial condition reads
		$$\mathbf{f}(0,u)=\mathbf{f}^0.$$
	\end{definition}
	\noindent Then, the following analytical result holds for the initial value problem \eqref{ivp}.

	\begin{theorem}\label{th1}
		Let the assumptions \eqref{assp}, \eqref{assp1}, \eqref{asspcon1}, \eqref{asspcon2} and \eqref{assp4} hold true. Then, there exists a unique local solution
		$$\mathbf{f}(t,u)\in \left(C\left([0,t_0]; L^1(D_u)\right)\right)^n$$ 
		of the initial value problem \eqref{ivp}, for some $t_0>0$. {Moreover, this function is positive, i.e. for all $i \in \{1, 2, \dots, n\}$, $f_i(t,u)\geq 0$, for all $t \in [0,\, t_0]$ and $u \in D_u$.}
		
		\begin{proof}
			Let $\mathbf{f}$ and $\mathbf{g}$ be two functions in $X=(L^1(D_u))^n$. In particular, there exists a constant $c>0$ such that
			\begin{align*}
				\|\mathbf{f}\|_1&\leq c,\\
				\|\mathbf{g}\|_1&\leq c.
			\end{align*}
			First, we consider
			$$\|\mathbf{\Sigma}[\mathbf{f}]-\mathbf{\Sigma}[\mathbf{g}]\|_1.$$
			Specifically, for $i \in \{1, 2, \dots, n\}$, by \eqref{assp4}, one has
			\begin{equation}\label{eqth11}\begin{split}
					|\Sigma_i[\mathbf{f}]-\Sigma_i[\mathbf{f}]|&\leq \sum_{h,k=1}^n\int_{D_u \times D_u}\left|\eta_{hk}(u_*,u^*)\mathcal{A}_{hk}^i(u,u_*,u^*)\left(f_h(t,u_*)f_k(t,u^*)-g_h(t,u_*)g_k(t,u^*)\right)\right|\, du_*du^*\\&
					\leq \eta\sum_{h,k=1}^n\int_{D_u \times D_u}\left|\mathcal{A}_{hk}^i(u,u_*,u^*)\left(f_h(t,u_*)f_k(t,u^*)-g_h(t,u_*)g_k(t,u^*)\right)\right|\, du_*du^*.
			\end{split}\end{equation}
			{Let integrate the \eqref{eqth11} over $D_u$, and sum on $i =1,2, \dots, n$. Since $\mathcal{A}_{hk}^i(u,u_*,u^*)$ is a probability and due to the assumption \eqref{assp}, one has 
			\begin{equation}\label{eqth122}\begin{split}
				\sum_{i=1}^n\int_{D_u}|\Sigma_i[\mathbf{f}]-\Sigma_i[\mathbf{f}]|\, du &\leq \eta\sum_{i,h,k=1}^n \int_{D_u^3}\mathcal{A}_{hk}^i(u,u_*,u^*)\left|\left(f_h(t,u_*)f_k(t,u^*)-g_h(t,u_*)g_k(t,u^*)\right)\right|\, du_*du^*du\\&
				\leq \eta\sum_{h,k=1}^n \int_{D_u^3}\left|\left(f_h(t,u_*)f_k(t,u^*)-g_h(t,u_*)g_k(t,u^*)\right)\right|\, du_*du^*du.
			\end{split}\end{equation}
			By summing and subtracting $f_h(t,u_*)g_k(t,u^*)$ in the right-hand side of the \eqref{eqth122}, some calculations show that
			\begin{equation}\label{eqth12}\begin{split}
					\sum_{i=1}^n\int_{D_u}|\Sigma_i&[\mathbf{f}]-\Sigma_i[\mathbf{f}]|\, du \leq \\&
					\leq \eta \sum_{h,k=1}^n \int_{D_u \times D_u} \left|\left[f_h(t,u_*)\left(f_k(t,u^*)-g_k(t,u^*)\right)+g_k(t,u^*)\left(f_h(t,u_*)-g_h(t,u_*)\right)\right]\right|\, du_*du^*\\&
					\leq \eta \sum_{h,k=1}^n \int_{D_u}f_h(t,u_*)\, du_* \int_{D_u}\left|f_k(t,u^*)-g_k(t,u^*)\right|\, du^*+\\&
					+\eta \sum_{h,k=1}^n \int_{D_u}g_k(t,u^*)\, du^*\int_{D_u}\left|f_h(t,u_*)-g_h(t,u_*)\right|\, du^*.
			\end{split}\end{equation}}
			Finally, \eqref{eqth12} can be rewritten as			
			{\begin{equation}\label{eqth13}\begin{split}
					\|\mathbf{\Sigma}[\mathbf{f}]-\mathbf{\Sigma}[\mathbf{g}]\|_1&\leq \eta \|\mathbf{f}\|_1\|\mathbf{f}-\mathbf{g}\|_1+\eta\|\mathbf{g}\|_1\|\mathbf{f}-\mathbf{g}\|_1\\&
					\leq 2\eta c\|\mathbf{f}-\mathbf{g}\|_1.
			\end{split}\end{equation}}
			Let us now consider the operator $\Omega_i[\mathbf{f}]$. In particular, for $i \in \{1, 2, \dots, n\}$, one has			
			\begin{equation}\label{eqth14}\begin{split}
					\left|f_i(t,u)\Omega_i[\mathbf{f}]-g_i(t,u)\Omega_i[\mathbf{g}]\right|&\leq\sum_{k=1}^n\int_{D_u}\left|\eta_{ik}(u,u^*)\left(f_i(t,u)f_k(t,u^*)-g_i(t,u)g_k(t,u^*)\right)\right|\, du^*\\&
					+\sum_{h=1, h \neq i}\int_{D_u}\left|\Gamma_{ih}(t)\left(f_i(t,u)f_h(t,u^*)-g_i(t,u)g_h(t,u^*)\right)\right|\, du^*\\&
					+\left|\Theta_i(t)\left(f_i(t,u)-g_i(t,u)\right)\right|.
			\end{split}\end{equation}
			{In light of the above assumptions \eqref{asspcon1} and \eqref{asspcon2}, there exist $\Gamma, \Theta>0$ such that
			\begin{align}
				\Gamma_{ih}(t)&\leq \Gamma, \qquad \forall i,h \in \{1, 2, \dots, n\},\label{assp2}\\
				\Theta_i(t)&\leq \Theta, \qquad \forall i \in\{1, 2, \dots, n\}\label{assp3}.
			\end{align}}
			Therefore, by using the \eqref{assp2}, \eqref{assp3} and \eqref{assp4}, from the \eqref{eqth14}, straightforward calculations show			
			\begin{equation}\label{eqth15}\begin{split}
					\left|f_i(t,u)\Omega_i[\mathbf{f}]-g_i(t,u)\Omega_i[\mathbf{g}]\right|&\leq \eta \sum_{k=1}^n\int_{D_u}|f_i(t,u)||f_k(t,u^*)-g_k(t,u^*)|\, du^*\\&
					+\eta\sum_{k=1}^n\int_{D_u}|f_i(t,u)-g_i(t,u)||g_k(t,u^*)|\, du^*\\&
					+\Gamma\sum_{h=1, h \neq i}\int_{D_u}|f_i(t,u)||f_h(t,u^*)-g_h(t,u^*)|\, du^*\\&
					+\Gamma\sum_{h=1, h \neq i}\int_{D_u}|f_i(t,u)-g_i(t,u)||g_h(t,u^*)|\, du^*\\&
					+\Theta |f_i(t,u)-g_i(t,u)|.
			\end{split}\end{equation}
			By integrating on $D_u$ and summing on $i \in \{1, 2, \dots, n\}$, \eqref{eqth15} becomes			
			\begin{equation}\label{eqth16}
				\sum_{i=1}^n\int_{D_u}	\left|f_i(t,u)\Omega_i[\mathbf{f}]-g_i(t,u)\Omega_i[\mathbf{g}]\right|\, du\leq 2\eta c\|\mathbf{f}-\mathbf{g}\|_1+2\Gamma c \|\mathbf{f}-\mathbf{g}\|_1+\Theta\|\mathbf{f}-\mathbf{g}\|_1.
			\end{equation}
			Finally, from the inequalities \eqref{eqth13} and \eqref{eqth16}, we conclude that			
			\begin{equation}\label{eqth17}
				\|\mathbf{C}[\mathbf{f}]-\mathbf{C}[\mathbf{g}]\|_1\leq 4\eta c \|\mathbf{f}-\mathbf{g}\|_1+2\Gamma c\|\mathbf{f}-\mathbf{g}\|_1+\Theta \|\mathbf{f}-\mathbf{g}\|_1.
			\end{equation}
			Let
			$$L:=4\eta c+2\Gamma c+\Theta,$$
			where the constant $c$ depends on $\|\mathbf{f}_0\|_1$. Then, the \eqref{eqth17} reads
			$$\|(\mathbf{C}[\mathbf{f}])-(\mathbf{C}[\mathbf{g}])\|_1\leq L\|\mathbf{f}-\mathbf{g}\|_1.$$
			{Therefore, the operator $\mathbf{C}[\mathbf{f}](t)$ is Lipschitz. Then, by using an extension of Picard-Lindel\"of Theorem on Banach spaces (see Chapter $1$ \cite{deimling2006ordinary} for details), since the initial data $\mathbf{f}^0(u)$ is chosen according to \eqref{assp1}, there exists a unique local solution of the initial value problem \eqref{ivp}. Specifically, there exists a unique solution $\mathbf{f}\in \left(C\left([0,t_0]; L^1(D_u)\right)\right)^n$ of the initial value problem \eqref{ivp} in the sense of the above Definition \ref{def}, for some $t_0>0$.}
			
			{To prove the positivity of the solution, we assume the compact form of the equation \eqref{eqfkin} according to the Definition \ref{def}, i.e.
			\begin{equation}\label{eqcomp}
			\frac{d}{d\,t}f_i(t,u)=\Sigma_i[\mathbf{f}](t,u)-f_i(t,u)\Omega_i[\mathbf{f}](t,u), \qquad i \in \{1, 2, \dots, n\},
			\end{equation}
			where the operators $\Sigma_i[\mathbf{f}](t,u)$ and $\Omega_i[\mathbf{f}](t,u)$ are defined, for all $i \in \{1, 2, \dots, n\}$, in \eqref{opersigma} and \eqref{operomega}, respectively. The equation \eqref{eqcomp}, for $i \in \{1, 2, \dots, n\}$ can be seen as a first-order nonhomogeneous ordinary differential equation in the unknown $f_i(t,u)$. Then, according to the classical theory of ordinary differential equations, the \eqref{eqcomp} writes, for $t \in [0,\, t_0]$ and $i \in \{1, 2, \dots, n\}$, as
			\begin{equation}\label{eqth18}
				f_i(t,u)=f_i^0\exp\left(-\int_0^t\Omega_i[\mathbf{f}](\tau,u)\, d\tau\right)+\int_0^t \Sigma_i[\mathbf{f}](\tau,u)\exp\left(-\int_{\tau}^t\Omega_i[\mathbf{f}](s,u)\,ds\right)\, d\tau.
			\end{equation}}
			The positivity of exponential function and operator $\Sigma_i[\mathbf{f}]$, for all $i \in \{1, 2, \dots, n\}$, ensures the positivity of $\mathbf{f}(t,u)$, solution of the initial value problem \eqref{ivp}. This concludes the proof.
			
		\end{proof}
	\end{theorem}
	
	{
	\begin{remark}
		The assumptions of the above Theorem $1$ allow to obtain a Lipschitz operator that proves the existence of a unique and positive solution, at least locally in time for the initial value problem \eqref{ivp}. It is worth pointing out that the assumption \textbf{H2} strictly regards the external force field \eqref{forshap2}. Indeed, it allows to keep bounded the solution, even if for a certain time-interval $[0,\, t_0]$. Analogously, it is worth noting that the analytical shape of the external force field \eqref{forshap2}, along with positivity of coefficients $\alpha_i$, for all $i \in \{1, 2, \dots, n\}$, preserves the positivity of solution, once a positive initial data $\mathbf{f}^0(u)$ has been assigned, regardless, in this case, of the assumption \textbf{H2}.  
	\end{remark}}

	\begin{remark}\label{rem1}
		Theorem \ref{th1} does not guarantee the global existence of solution of the initial value problem \eqref{ivp}. Indeed, $\mathbf{f}(t,u)$ is defined in a time interval $[0,\, t_0]$, where $t_0>0$ generally depends on initial data $\mathbf{f}^0$ and parameters of the system. Indeed, in a nonconservative kinetic framework blow-up phenomena may occur, \cite{arlotti1996qualitative}. In order to gain a possible condition for the global existence of solution of the initial value problem \eqref{ivp}, let integrate and sum on $i\in \{1, 2, \dots, n\}$, the kinetic equation \eqref{eqfkin}. Then, by using the assumption \eqref{assp}, one has
		\begin{equation}\label{eqcons}\begin{split}
				\rho(t)&=\rho(0)+\sum_{i=1}^n\sum_{h=1, h \neq i}^{n}\int_0^t\Gamma_{i h}(\tau)\, \rho_i(\tau)\rho_h(\tau)\, d\tau+ \int_0^t\sum_{i=1}^n \rho_i(\tau)\Theta_i (\tau)\, d\tau+|D_u|\int_0^t \sum_{i=1}^n\alpha_i \,d\tau\\&
				=\rho(0)+\sum_{i=1}^{n}\int_0^t\left(\sum_{h=1, h \neq i}^{n}\Gamma_{i h}(\tau)\, \rho_i(\tau)\rho_h(\tau)+\rho_i(\tau)\Theta_i (\tau)+|D_u|\sum_{i=1}^n\alpha_i \right)d\tau,
		\end{split}\end{equation}
		where $|D_u|$ is the measure of set $D_u$. Generally, the boundedness of $\mathbf{f}(t,u)$ for all $t>0$ ensures the global existence in time of solution. Therefore, if 
		\begin{equation}\label{eqbounded}
			\int_0^t\left(\sum_{h=1, h \neq i}^{n}\Gamma_{i h}(\tau)\, \rho_i(\tau)\rho_h(\tau)+\rho_i(\tau)\Theta_i (\tau)+|D_u|\sum_{i=1}^n\alpha_i \right)d\tau< +\infty, \qquad \forall t>0,
		\end{equation}
		then the local solution can be extended globally in time.
		
		The boundedness condition \eqref{eqbounded} may be ensured by time-decaying or periodic shape external force fields. Nevertheless, this paper does not aim to derive a final result in this direction, even with respect to its modeling purposes.
		
	\end{remark}

	\section{Application to autoimmune response treatment}\label{secappl}

	As already mentioned, kinetic theory for interacting agents can be applied to several frameworks, in particular, biology and cellular dynamics. Indeed, integral operators of type \eqref{kineqGi} and \eqref{kineqLi} can suitably reproduce interactions among different populations of cells, while the activity variable $u$ represents the level of activation of a cell in performing its own task. In this section, we apply it to the specific case of autoimmunity in the presence of therapy protocols.

	\subsection{Kinetic description}\label{subsectkin}
	For our purposes, we implement the kinetic framework \eqref{eqfkin}. Firstly, each population is described by a distribution function, depending on time $t\geq 0$ and activity variable $u$; the latter acquires its values in $D_u=[0,\,1]$. Specifically, the following $4$ functional subsystems, along with their activity variable, are considered
	
	\begin{itemize}
		\item[-] Self-antigen presenting cells (SAPCs), the activity is the ability to activate immune cells;
		\item[-] Naive T-cells (NBCs), their activity represents the capacity to recognize SAPCs antigen;
		\item[-] Memory self-reactive B cells (MSBCs), the activity variable accounts for the level of cytokines production;
		\item[-] Tregs, the activity is the efficiency in detecting and suppressing SAPCs and MSBCs.
	\end{itemize}
	Then, each population is described by its distribution function $f_I(t,u)$, where the subindex $I$ indicates the population. More precisely letters $A$, $N$, $M$, $T$ indicate SAPCs, NBCs, MSBCs, and Tregs, respectively. For our purposes, we take the SAPCs population as a background, supposing that it is constant in time. On the other hand, we take into account the fact that MSBCs may also act as SAPCs \cite{rodriguez2005b} participating in B-cells differentiation \cite{vazquez2015b} or stimulating CD4$^+$ T-cells expansion \cite{constant1999b}.
	
	In view of the application of this paper, the system \eqref{eqfkin} writes
		{
	\begin{align}\label{SistKinA}
		\frac{\partial}{\partial\,t} f_N(t,u)=& \, -\eta_{NA}\,  f_N(t,u)  \int_0^1 f_A(u^*) d u^*- \eta_{NM}\, f_N(t,u)  \int_0^1 f_M(t,u^*) d u^*	\\[2mm]
		\nonumber &\quad+
		\Theta_N(t) f_N(t,u)
		\\[2mm]
		\label{SistKinM}
		\frac{\partial}{\partial\,t} f_M(t,u)=& \, \eta_{NA}\, \int_0^1\int_0^1 f_A(u^*)\,f_N(t,u_*)\,d u^*\,du_*+ \eta_{NM}\, \int_0^1\int_0^1 f_M(t,u^*)\,f_N(t,u_*)\,d u^*\,du_*\nonumber
		\\[2mm]
		 &\quad+\Gamma_{MT}(t) f_M(t,u)\int_0^1 f_T(t,u^*)\,d u^* + \Theta_M(t)  f_M(t,u) +\alpha_M\\[2mm]
		\label{SistKinT}
		\frac{\partial}{\partial\,t} f_T(t,u)= & \,\Gamma_{TM}(t)f_T(t,u)\int_0^1 f_M(t,u^*)\,d u^* + \Theta_T(t) f_T(t,u)+\alpha_T.
	\end{align}
	
       In the system above we have included, in relation to the kinetic model previously described, the \emph{conservative dynamics}  that involves the B-cell population. In particular, an NBC becomes an MSBC interacting with either SAPCs (with constant interaction rate $\eta_{NA}>0$) or another MSBC  (with constant interaction rate $\eta_{NM}>0$). In both cases, the transition probabilities $\mathcal A_{NA}^M$ and  $\mathcal A_{NM}^M$ are assumed uniformly equal to $1$. This set definitely determines both the gain-term operator \eqref{kineqGi} in equation \eqref{SistKinA} and the loss-term operator \eqref{kineqLi} in \eqref{SistKinM}.
       
The components of the external force field \eqref{forshap}, each related to a specific population, include different \emph{nonconservative processes}.
\subsubsection*{Nonconservative processes for NBCs}
For NBCs the nonconservative processes in  \eqref{SistKinA} consist of 	\begin{equation}\label{TetaN}\Theta_{N}(t)=\tau_N-\sigma_N\,R_t(t), \end{equation}
where we have included the following terms:
\begin{itemize}
\item[-]  Natural proliferation (due to a lack of negative selection) with constant rate $\tau_N>0$.
\item[-] Medical suppression using Rituximab treatment. In order to reproduce therapy protocols reported in literature \cite{gottenberg2005tolerance}, that consist of weekly infusions of Rituximab for 2/4 weeks, we choose a function that suitably models the treatment, but that also fulfills analytical properties stated in the previous sections; this is given by
\begin{equation}\label{RituFun}
R_t(t)=\frac{\gamma_{Rt}}{2}\,\left({\sin\left(2\,\pi\,t - \frac12\,t\right) + 1}\right)\,\mathbb 1_{U_1},\end{equation}
with $\mathbb 1_{U_1}$ is the indicator function on the set $U_1=[0,1]\cup[7,8]\cup[14,15]\cup[21,22]$, and $\gamma_{Rt}>0$ stands for the dose. Then, the term $\sigma_N>0$ is the response rate of NBCs to therapy.
\end{itemize}

\subsubsection*{Nonconservative processes for MSBCs}			
For MSBCs, instead, the external force field in \eqref{SistKinM} includes:
\begin{itemize}
	\item The term
\begin{equation}
	\label{GammaMT}\Gamma_{MT}(t)=-\sigma_{MT}\,IL2(t),\end{equation}		
	describing
 \emph{destructive interactions} between  MSBCs and Tregs induced by  IL-2 therapy;  also in this case, we refer to medical studies \cite{koreth2011interleukin} where the IL-2 interleukin intake is a daily dose for 8 weeks, thus it is represented by the term 
\begin{equation}\label{ILFun}
	IL2(t)=\frac{\gamma_{IL2}}{2}\,\left({\sin\left(2\,\pi\,t - \frac12\,t\right) + 1}\right)\,\mathbb 1_{U_2},\end{equation} with $U_2=[0,56]$, , $\gamma_{IL2}>0$ stands for the dose.. In addition, $\sigma_{MT}>0$ is the response rate of Tregs to the therapy in increasing their suppressive function.
\item 
Further \emph{nonconservative processes} given by the term		\begin{equation}\label{TetaM}\Theta_{M}(t)=-d_M-\sigma_M\,R_t(t),\end{equation} where we have the following:
\begin{itemize}
\item[-] natural decay with constant rate $d_M>0$,
\item[-]  medical suppression using Rituximab treatment, being $\sigma_M>0$ the response rate to the therapy and $R_t(t)$ defined in \eqref{RituFun}.
\end{itemize}
\item The tendency of MSBCs to relax toward a minimum value. This is represented in the equation \eqref{SistKinM} by the term $\alpha_M=d_M\,\bar M$, $\bar M>0$.
\end{itemize}		

\subsubsection*{Nonconservative processes for Tregs}
Finally, for Tregs, the right-hand side of $\eqref{SistKinT}$ describes:
\begin{itemize}
\item \emph{Proliferative interactions} with  MSBCs occurring at constant rate $\Gamma_{TM}(t)=\beta_{TM}$; 
\item Other \emph{nonconservative} terms giving
\begin{equation}\label{TetaT}\Theta_{T}(t)=-d_T+\sigma_T\,IL2(t),\end{equation}
with:
\begin{itemize}
\item[-] natural decay with constant rate $d_T>0$;
\item[-] natural proliferation is stimulated by low doses of $IL-2$,
		where $\sigma_T>0$ is the Tregs response rate to therapy and $IL2(t)$ is given in \eqref{ILFun}.
\end{itemize}
		\item The fact that Tregs relax toward a minimum value; this is represented in the equation \eqref{SistKinT} by the term $\alpha_T=d_T\,\bar T$, $\bar T>0$.
	\end{itemize}

\noindent Bearing all above in mind, the system  \eqref{SistKinA}-\eqref{SistKinT} can be rewritten explicitly as follows:
	\begin{align}\label{SistKinAe}
	\frac{\partial}{\partial\,t} f_N(t,u)=& \, -\eta_{NA}\,  f_N(t,u)  \int_0^1 f_A(u^*) d u^*- \eta_{NM}\, f_N(t,u)  \int_0^1 f_M(t,u^*) d u^*	\\[2mm]
	\nonumber &\quad+
	\left[\tau_N-\sigma_N\,\left(\frac{\gamma_{Rt}}{2}\,\left({\sin\left(2\,\pi\,t - \frac12\,t\right) + 1}\right)\,\mathbb 1_{U_1}\right)\right] f_N(t,u)
	\\[2mm]
	\label{SistKinMe}
	\frac{\partial}{\partial\,t} f_M(t,u)=& \, \eta_{NA}\, \int_0^1\int_0^1 f_A(u^*)\,f_N(t,u_*)\,d u^*\,du_*\nonumber
	\\[2mm]
	&\quad+ \eta_{NM}\, \int_0^1\int_0^1 f_M(t,u^*)\,f_N(t,u_*)\,d u^*\,du_*\nonumber
	\\[2mm]
	&\quad-\sigma_{MT}\,\left[\frac{\gamma_{IL2}}{2}\,\left({\sin\left(2\,\pi\,t - \frac12\,t\right) + 1}\right)\,\mathbb 1_{U_2}\right]\, f_M(t,u)\int_0^1 f_T(t,u^*)\,d u^* \nonumber
	\\[2mm]
	&\quad-\left[d_M+\sigma_M \left(\frac{\gamma_{R_t}}{2}\,\left({\sin\left(2\,\pi\,t - \frac12\,t\right) + 1}\right)\,\mathbb 1_{U_1}\right)\right] f_M(t,u) +\alpha_M\\[2mm]
	\label{SistKinTe}
	\frac{\partial}{\partial\,t} f_T(t,u)= & \,\beta_{TM}f_T(t,u)\int_0^1 f_M(t,u^*)\,d u^* \nonumber
	\\[2mm]
	&\quad +\left[-d_T+\sigma_T\left(\frac{\gamma_{IL2}}{2}\,\left({\sin\left(2\,\pi\,t - \frac12\,t\right) + 1}\right)\,\mathbb 1_{U_2}\right)\right]\,f_T(t,u)+\alpha_T.
\end{align}

 }
	Assigned a suitable initial data 
	\begin{equation}
		\label{DatoKin}\mathbf{f}^0(u)=\left(f^0_A(u),\, f^0_N(u),\, f^0_M(u),\, f^0_T(u)\right),\end{equation}
	the results of Theorem \ref{th1} and considerations of Remark \ref{rem1} ensure that the system \eqref{SistKinAe}-\eqref{SistKinTe} has a unique positive solution, at least locally in time. Moreover, due to the choice of periodic parameters for the external force field, this may ensure the boundedness of the solution, with its consequent global extension in $[0,\, + \infty]$. At the moment, evidences of such behavior are provided numerically.
	
	\subsection{Macroscopic study}\label{Subsecmacro}
	Now, our purpose is the study of evolution in time of macroscopic densities of NBCs, MBCs, and Tregs, that are
	$$I(t)=\displaystyle\int_0^1f_I(t,u)\,du, \qquad I\in \{N,\,M,\,T\},$$ while, for APCs we define 
	$$\widetilde A=\displaystyle\int_0^1f_A^0(u)\,du.$$
	Therefore, by integrating the equations \eqref{SistKinAe} -- \eqref{SistKinTe} with respect to the activity variable $u$, one has
	{
	\begin{align}\label{SistComA}
		\frac{d\,N(t)}{d\,t} =&  - N(t)\left(\eta_{NA}\,\widetilde A+\eta_{NM}\,M(t)\right)+\left(\tau_N-\sigma_N\,R_t(t)\right)\,N(t)\\[2mm]
		\label{SistComM}
		\frac{d\,M(t)}{d\,t} =&\,N(t)\left( \eta_{NA}\,\widetilde A+\eta_{NM}\, M(t)\right) \nonumber
		\\[2mm]
		&\quad
		-\sigma_{MT}IL2(t)\,  M(t) T(t)-\left(d_M+\sigma_M\,R_t(t)\right)\, M(t) +\alpha_M\\[2mm]
		\label{SistComT}
		\frac{d\,T(t)}{d\,t} =&\,\beta_{TM}\,M(t)\,T(t)
	+\left(-d_T+\sigma_T\,IL2(t)\right)
		\, T(t)+\alpha_T,
	\end{align}
    where functions $R_t(t)$ and $IL2(t)$ are defined in \eqref{RituFun} and \eqref{ILFun}, respectively. }
	The initial data is, therefore, 
	\begin{equation}\label{DatoMac}\boldsymbol{\rho}^0(u)=\left(N^0(u),\, M^0(u),\, T^0(u)\right),\end{equation} where $ I^0$ represents the integral over $u$ of the corresponding initial datum $f_I^0(u)$, as given in \eqref{DatoKin}.

	\noindent Moreover, the set
	{\begin{equation}
		\mathcal E =\left\{(N(t),\,M(t),\,T(t)) \in\mathbb R^3 : (\widetilde A\geq 0,\,N(t)\geq 0,\,M(t)\geq 0,\,T(t)\geq 0\right\}
	\end{equation}
}
	represents the set of all biologically significant solutions for \eqref{SistComA} -- \eqref{SistComT}, with initial data \eqref{DatoMac}.
	
	Generally, is not possible to find an explicit solution of the system \eqref{SistComA}-\eqref{SistComT}. Furthermore, it is nonautonomous. Therefore, the system without treatment (which is autonomous) is considered at first, and, afterward, the treatment terms are added. The system \eqref{SistComA} -- \eqref{SistComT}, hence, writes
	\begin{align}\label{SistNoTeA}
		\frac{d\,N(t)}{d\,t} =&  - N(t)( \eta_{NA}\,\widetilde A+\eta_{NM}\, M(t))+\tau_N\,N(t)\\[2mm]
		\label{SistMoTeM}
		\frac{d\,M(t)}{d\,t} =&N(t)( \eta_{NA}\,\widetilde A+\eta_{NM}\, M(t))-d_M\, M(t)+\alpha_M \\[2mm]
		\label{SistNoTeT}
		\frac{d\,T(t)}{d\,t} =&\beta_{TM}\,M(t)\,T(t) -d_T\, T(t)+\alpha_T.
	\end{align}
	Equating the right-hand sides of equations \eqref{SistNoTeA}-\eqref{SistNoTeT} to zero, two steady states are found, i.e. a boundary one
	\begin{equation}\label{eq1}
		E_1=\left(0,\,\bar M,\,\frac{\alpha_T}{d_T- \bar M\,\beta_{TM}  }
		\right),
	\end{equation}
	that belongs to $\mathcal E$  only if the following condition is satisfied
	\begin{equation}\label{Eq1Ex}
		\Delta_1:=d_T- \bar M\,\beta_{TM}>0;
	\end{equation}
	and
	\begin{equation}\label{eq2}
		E_2=\left(\frac{ d_M\,\left(\tau_N-\bar A\, - \bar M\,\eta_{NM}\,\right)}{\eta_{NM}\,\,\tau_N}
		,\, \frac{\tau_N - \bar A\,}{\eta_{NM}\,}\,, 
		\frac{\alpha_T\,\eta_{NM}\,}{  d_T\,\eta_{NM}\, + \beta_{TM}\,\left(\bar A\,-\tau_N\right)}
		\right),
	\end{equation}
	{ where we have assumed that $\bar A:=\widetilde{A}\,\eta_{NA}$. }
	The equilibrium $E_2$ is internal to $\mathcal E$  only if the following conditions are satisfied
	\begin{equation}\label{Eq2Ex}
		\Delta_3:=\Delta_2+\bar M\,\eta_{NM}\,<0,\quad \Delta_4:=\beta_{TM}\,\Delta_2+d_T\,\eta_{NM}\,>0,\mbox{ with } \Delta_2:=\bar A\,-\tau_N.
	\end{equation}
	Concerning the stability of equilibria, the following result holds.
	\begin{theorem}\label{th2}
		Let $E_1,\,  E_2$, whose analytical expression is provided in \eqref{eq1} -- \eqref{eq2}, be the equilibria of system \eqref{SistNoTeA}-\eqref{SistNoTeT}. Let condition \eqref{Eq1Ex} hold, we have the following cases
		\begin{enumerate}
			\item If $\Delta_2>0$ or $\Delta_2<0$ and at the same time $\Delta_3>0$, only equilibrium  $E_1$ exists and it is locally asymptotically stable.
			\item If $\Delta_3<0$ and $\Delta_4>0$, equilibrium $E_1$ is unstable and equilibrium $E_2$  is locally asymptotically stable.
			\item If $\Delta_3<0$ and $\Delta_4<0$, only equilibrium  $E_1$ exists and it is locally unstable.
		\end{enumerate}
	\end{theorem}
	
	\proof 
	The results on existence come straightforward from \eqref{Eq1Ex} and \eqref{Eq2Ex}. To perform stability analysis towards equilibria, let us consider the Jacobian matrix associated with the system \eqref{SistNoTeA} -- \eqref{SistNoTeT}, that is 
	\begin{equation}{\mathbf J}=
		\left(\begin{array}{ccc}
			-\bar A - M\,\eta_{NM}\, + \tau_N & -N\,\eta_{NM}\, & 0 \\
			&&\\
			\bar A + M\,\eta_{NM}\, & -d_{M} + N\,\eta_{NM}\, & 0 \\
			&&\\
			0 & T\,\beta_{TM} & -d_T + M\,\beta_{TM}
		\end{array}\right).
	\end{equation}
	The Jacobian $\mathbf J$ evaluated in equilibrium $E_1$ given in \eqref{eq1} reads
	\begin{equation}{\mathbf J}(E_1)=
		\left(
		\begin{array}{ccc}
			-\bar A -{ \bar M \,\eta_{NM}\,} + \tau_N & 0 & 0 \\
			&&\\
		\bar A +{ \bar M \,\eta_{NM}\,} & - d_M & 0 \\
			&&\\
			0 & \dfrac{\alpha_T\, \beta_{TM}}{d_T - \bar M \beta_{TM}} & -d_T +{\bar M\,\beta_{TM}}
		\end{array}
		\right).
	\end{equation}
	The characteristic polynomial of ${\mathbf J}(E_1)$ turns out to be
	\begin{equation}
		P_1(\lambda)=\lambda^3 +A_1\, \lambda^2 +B_1 \, \lambda + C_1,
	\end{equation}
	with coefficients
	$$
	\begin{aligned}
		A_1&=&{{\Delta_1  + \Delta_3+ {d_M}}},\quad
		B_1&=& \Delta_1\,\Delta_3+d_M\,(\Delta_1+\Delta_3),\quad
		C_1&=&{{\Delta_1 \,\Delta_3\,}}{d_M}.
	\end{aligned}
	$$
	According to the Routh-Hourwitz criterion \cite{gantmacher1959applications}, all the roots of polynomial $P_1$ are negative or have negative real part if $A_1>0$, $C_1>0$ and $A_1\,B_1>C_1$. By simple algebra we get
	$$A_1\,B_1-C_1={{(\Delta_1 + \Delta_3) (\Delta_1 + {d_M}) (\Delta_3 + {d_M})}}
	$$
	Thus we firstly observe that the condition $\Delta_3>0$ is necessary for the stability of $E_1$ and from expression \eqref{Eq2Ex} we can deduce the validity of points $1.$ and $3.$ of the Theorem. To prove point $2.$, we calculate the Jacobian $\mathbf J$ in equilibrium $E_2$, obtaining
	\begin{equation}{\mathbf J}(E_2)=
		\left(
		\begin{array}{ccc}
			0 & \dfrac{d_M \,\left( \bar A + \bar M \, \eta_{NM}\, -  \, \tau_N\right)}{\tau_N} & 0 \\
			&&\\
			\tau_N & -\dfrac{d_M \,\left( \bar A + \bar M \, \eta_{NM}\,\right)}{\tau_N} & 0 \\
			&&\\
			0 & \dfrac{\alpha_T \, \beta_{TM} \, \eta_{NM}\,}{\beta_{TM} (\bar A - \tau_N) + d_T \, \eta_{NM}\,} & -d_T + \dfrac{\beta_{TM} (-\bar A + \tau_N)}{\eta_{NM}\,}
		\end{array}
		\right).
	\end{equation}
	The characteristic polynomial of ${\mathbf J}(E_2)$ is
	\begin{equation}
		P_2(\lambda)=\lambda^3 +A_2\, \lambda^2 +B_2 \, \lambda + C_2,
	\end{equation}
	with coefficients
	$$
	\begin{aligned}
		&A_2=\frac{\Delta_4}{\eta_{NM}\,} + \frac{d_M\left( \bar A + \bar M \eta_{NM}\,\right)}{\tau_N},\quad
		B_2=d_M\,\left( \frac{\Delta_4\,\left( \bar A + \bar M \eta_{NM}\,\right)}{\eta_{NM}\, \tau_N} -\Delta_3\right),\\
		&C_2=-\frac{d_M\,\Delta_3 \Delta_4}{\eta_{NM}\,}.
	\end{aligned}
	$$
	Relying again on the Routh-Hurwitz criterion, since 
	$$A_2\,B_2-C_2=\frac{d_M\,\left( \bar A + \bar M\,\eta_{NM}\,\right) \left(\Delta_4\, \eta_{NM}\, d_M\,\left(\bar A + \bar M\,\eta_{NM}\,\right) +\tau_N\left( {\Delta_4}^2\,  - \Delta_3 \,{\eta_{NM}\,}^2\right)\, \right)}{{\eta_{NM}\,}^2\, {\tau_N}^2} 
	,$$
	we can state that, in the region of parameters where equilibrium $E_2$ exists, it is locally asymptotically stable.
	\endproof
	
	\noindent { Hereafter, for purely illustrative reasons, parameters of the model (omitting therapy) are set as follows
	\begin{align}
		\eta_{NA}\,=2,\quad\eta_{NM}\,=1,\quad \widetilde{A}=0.04,\quad
		d_M=1,\quad d_T=1,\quad\beta_{TM}=1,\quad \alpha_T=0.01.\label{Pars}
	\end{align}
	We point out that the value for the constant population $\tilde A$ is inspired by the values resulting in previous mathematical models for autoimmunity \cite{ramos2019kinetic,della2022mathematical}. On the other hand, the relatively low value assumed for $\alpha_T$ concerns the fact that numerous studies indicate Tregs deficits as a potential root cause for autoimmune conditions \cite{brusko2008human}.}

\begin{figure}[ht!]
		\centering
		\includegraphics[scale=0.4]{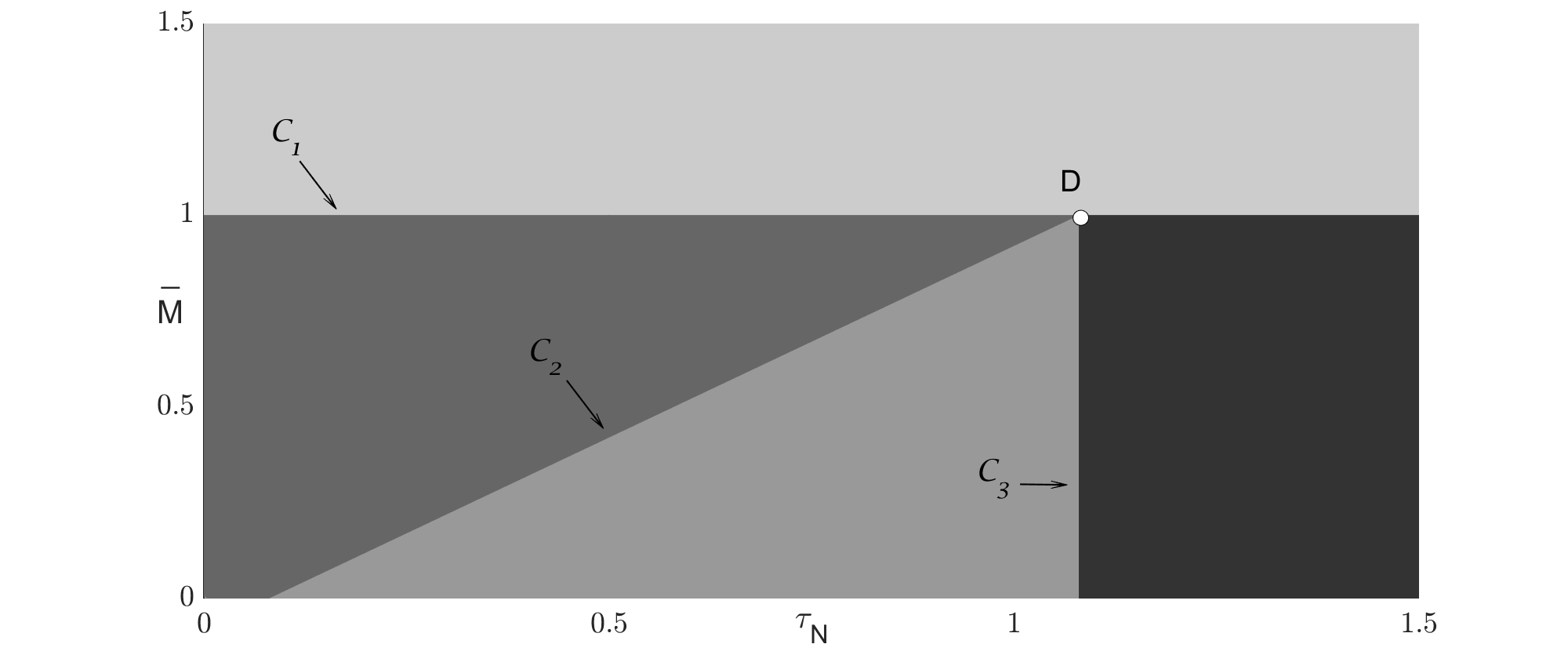}
		\caption{Bifurcation diagram in the space of parameters $(\tau_N, \bar M)$ for system \eqref{SistNoTeA}-\eqref{SistNoTeT}, assuming parameters as in \eqref{Pars}. Critical lines $\mathcal C_1$, $\mathcal C_2$, $\mathcal C_3$, given in \eqref{C1}, \eqref{C2} and \eqref{C3}, respectively, and point  $D: ( {{\tau}}_{N_c}, {{\bar M}_c})=(1.08,1)$ are also reported.}
		\label{fig1}
	\end{figure}

	We now discuss the existence and stability of equilibria in terms of parameters $\tau_N$ and $\bar M$. More precisely, let us take the parameters space $(\tau_N,\bar M)$ and individuate regions of parameters in which conditions stated in Theorem \ref{th2} are satisfied. Results are depicted in Figure \ref{fig1}. Condition \eqref{Eq1Ex} is satisfied for values below the horizontal line 
	\begin{equation}
		\label{C1}\mathcal C_1:\bar M=\frac{d_T}{\beta_{TM}},
	\end{equation}
	and equilibrium $E_1$ is stable for values to the left of the line 
	\begin{equation}
		\label{C2}\mathcal C_2:\bar M=\frac{\tau_{N}-\bar A}{\eta_{NM}\,},
	\end{equation}
	and unstable for values to the right.
	Moreover, equilibrium $E_2$ exists and it is stable in the area below $\mathcal C_1$ and between the curves  $\mathcal C_2$ and  $\mathcal C_3$, this last one given by  
	\begin{equation}
		\label{C3}\mathcal C_3: \tau_N=\frac{d_T\, \eta_{NM}\,}{\beta_{TM}}+\bar A.
	\end{equation}
	
	The critical point $D$ of intersections between $\mathcal C_1$, $\mathcal C_2$ and $\mathcal C_3$ reads 
	$$D:= ( {{\tau}}_{N_c}, {{\bar M}_c})=(1.08,1).$$

	\subsection{Numerical simulations}\label{Subsecnum}

	Here, some numerical simulations are performed, towards the kinetic model, by using { the standard  MATLAB tool \textsf{ode45}, built upon the explicit Runge-Kutta $(4,5)$ formula known as Dormand-Prince pair \cite{dormand1980family}}. Firstly, we aim to simulate the behavior of the solution in the nontreatment scenario (system \eqref{SistNoTeA}-\eqref{SistNoTeT}), showing stability results outlined analytically in the previous Subsection.  Furthermore, we want to inquire about the nonautonomous case \eqref{SistComA}-\eqref{SistComT}, in which therapy is included. Moreover, parameters are set such that the behavior of the solution, with respect to the nontreatment scenario, ensures the expected effects of the therapy on the autoimmune dynamics. 
	
	For the treatment-free case, assuming parameters as in \eqref{Pars}, referring to the bifurcation diagram in Figure \ref{fig1}, we take $\tau_N=0.5$, that provides the bifurcation value $\bar M=0.42$. We point out that, from a modeling point of view, only a choice of $\tau_N<{\tau_N}_c=1.08$ is meaningful. Thus,  we fix two different values for $\bar M$, more precisely $\bar M=0.1$ and  $\bar M=0.5$, that are one below and one above the line $\mathcal{C}_2$, respectively. Then, we want to test the effects of treatment in both cases, thus we  fix response parameters as follows
	\begin{align}
		\sigma_{N}=1,\quad\sigma_{MT}=1,&\quad\sigma_{M}=2,\quad\sigma_{T}=5.5.\label{ParsTer1}
	\end{align}
	For treatment doses instead, we refer to \cite{gottenberg2005tolerance} and \cite{koreth2011interleukin}, where the typical intakes are approximately  $375\,mg/m^{-2}$ for Rituximab and $0.06\,mg/m^{-2}$ for interleukin IL-2 (as maximal tolerated dose), then
	\begin{align}
		\gamma_{IL2}=0.06,&\quad\gamma_{Rt}=375.\label{ParsTer2}
	\end{align}
	Since in medical reports effects of therapy after a mean follow-up period of 8/12 months, we take as final time $t=200$ and we compare the treatment-free (system  \eqref{SistNoTeA} -- \eqref{SistNoTeT}) or treatment-delivery (system  \eqref{SistComA} -- \eqref{SistComT}) cases, still assuming other parameter as in \eqref{Pars}.

\begin{figure}[ht!]
		\centering
		\includegraphics[scale=0.55]{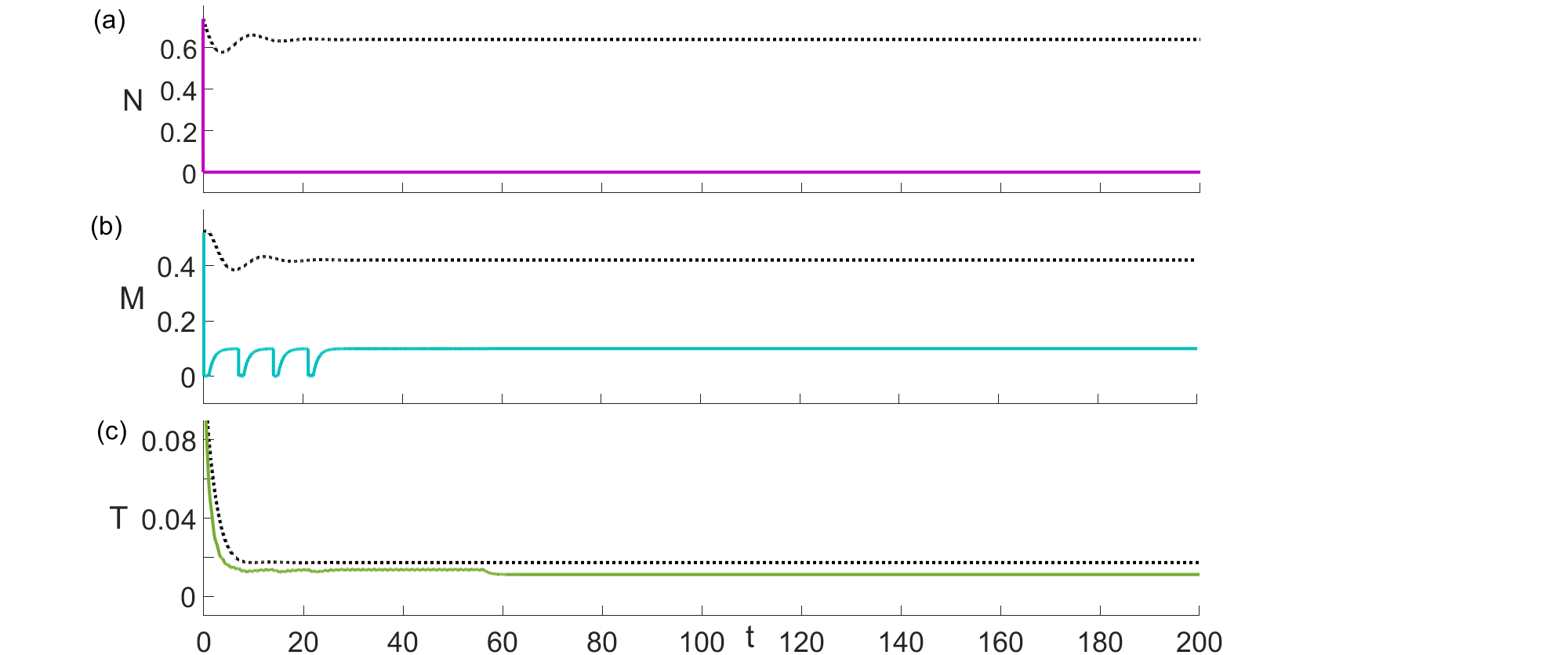}
		\caption{Numerical results for therapy-free system \eqref{SistNoTeA} -- \eqref{SistNoTeT} (dotted lines), compared with those for a complete system with therapy \eqref{SistComA} -- \eqref{SistComT}  (solid lines),  taking parameters as in \eqref{Pars}-\eqref{ParsTer1}-\eqref{ParsTer2}, $\tau_N=0.5$, $\bar M=0.1$. The behavior of macroscopic densities $N$, $M$, $T$ are reported in Panels (a), (b), (c), respectively.}
		\label{fig2}
	\end{figure}

	The first scenario ($\bar M= 0.1$) is the one in which equilibrium $E_2$ given in \eqref{eq2} exists and is stable, while equilibrium $E_1$ given in \eqref{eq1} is unstable, thus we perform simulations taking as initial data a perturbation of $E_2$. In this case, numeric values for equilibria are
	\begin{equation}
		\begin{aligned}\label{EqsSim1}
			E_1=&(0,\,0.1,\,0.11),\\
			E_2=&(0.64, 0.42, 0.017).
		\end{aligned}
	\end{equation}
	We may underline that, biologically, equilibrium $E_1$ represents a better condition, since the number of MSBCs is lower. Figure \ref{fig2} shows results obtained for the three involved populations. { Dotted lines represent the nontreatment scenario given by system \eqref{SistNoTeA}-\eqref{SistNoTeT}; in this case, we can observe the stability of equilibrium $E_2$ given in \eqref{EqsSim1}. By comparing it to the therapy scenario (solid lines) represented by the complete system \eqref{SistComA}-\eqref{SistComT}, first of all, we can see that the results are consistent with the theoretical results provided in the current paper, since the solution is positive and bounded in the time interval considered.
 In addition, from the modeling viewpoint, we can observe that treatment induces an immediate decrease of NBCs (Figure \ref{fig2} Panel (a)), while MSBCs (Figure \ref{fig2} Panel (b)) stabilize to a value that is lower than the equilibrium one in the treatment-free case. Finally, the Tregs population (Figure \ref{fig2} Panel (c)) exhibits lower values (due to less proliferation by interaction with MSBCs) with respect to the treatment-free case, with a slight increase during the IL-2 interleukin delivery, represented by the nonconstant terms in \eqref{GammaMT} and \eqref{TetaT}. We also report that, at initial stage, oscillations for populations of MSBCs and Tregs take place in correspondence of each intake of Rituximab, that is provided by the nonconstant terms in \eqref{TetaN} and \eqref{TetaM}.}
 
	Notably, the final values of populations for the treatment-delivery case coincide with those of equilibrium $E_1$. This means that the therapy can drive the system towards the second equilibrium which remains stable during the time interval considered. In particular, the number of B-cells is lower, and this is consistent with results from medical literature \cite{gottenberg2005tolerance}.
	
\begin{figure}[ht!]
		\centering
		\includegraphics[scale=0.55]{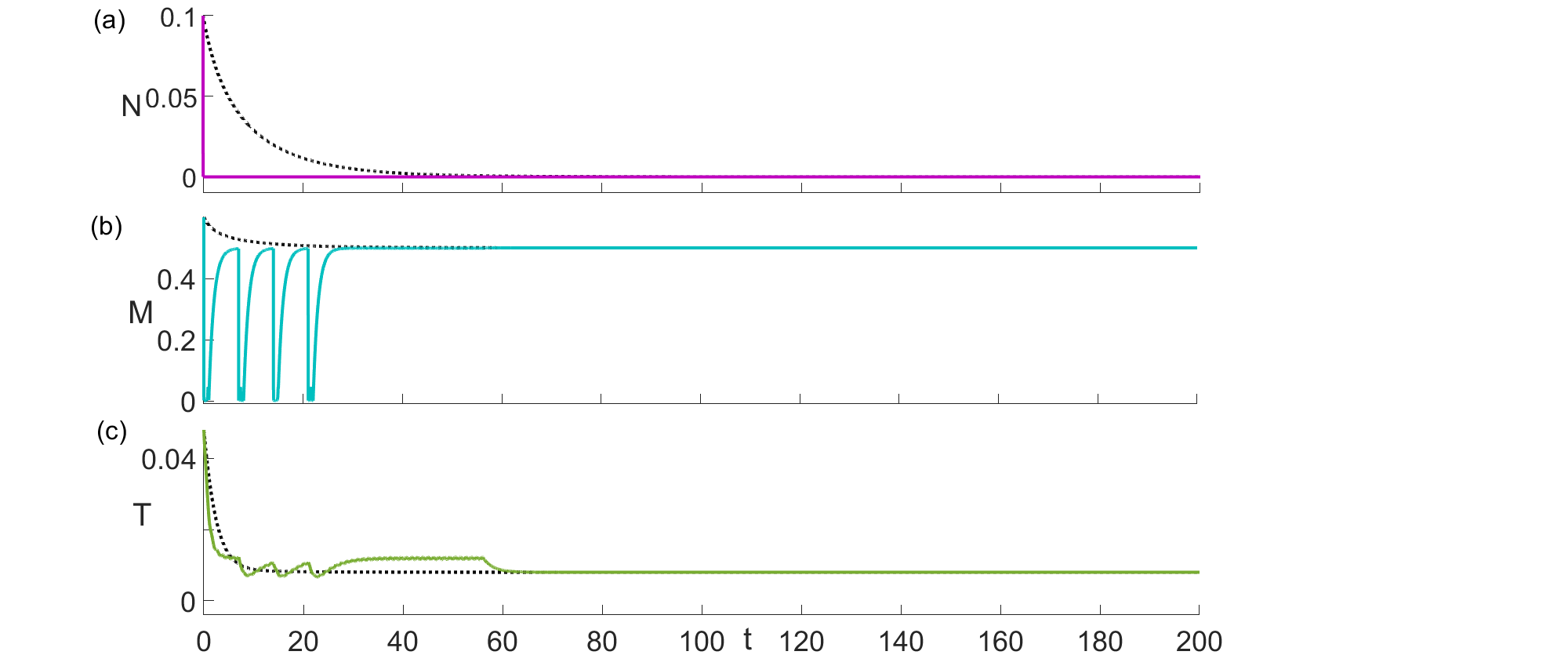}
		\caption{Numerical results for therapy-free system \eqref{SistNoTeA} -- \eqref{SistNoTeT} (dotted lines), compared with those for a complete system with therapy \eqref{SistComA} -- \eqref{SistComT}  (solid lines),  taking parameters as in \eqref{Pars}-\eqref{ParsTer1}-\eqref{ParsTer2}, $\tau_N=0.5$, $\bar M=0.5$. The behavior of macroscopic densities $N$, $M$, $T$ is reported in Panels (a), (b), (c), respectively.}
		\label{fig3}
	\end{figure}

	Now we consider the scenario with $\bar M=0.5$, in which only equilibrium $E_1$ given in \eqref{eq1} exists and is stable. In this case, we have
	\begin{equation}
		\begin{aligned}\label{EqE1Sim}
			E_1=&(0,\,0.5,\,0.02).
		\end{aligned}
	\end{equation}
	Taking as initial value a perturbation of $E_1$ and performing again simulations of systems  \eqref{SistNoTeA} -- \eqref{SistNoTeT} and \eqref{SistComA} -- \eqref{SistComT}, we may observe the impact of therapy in this case, reporting behavior of each population density in Figure \ref{fig3}.{  We report that, unlike the previously depicted case, the final values are now the same for both treatment-free (dotted lines) and treatment-delivery (solid lines) cases, and they coincide with equilibrium $E_1$ provided in \eqref{EqE1Sim}}. What can be pointed out is that, in the presence of therapy terms, NBCs reach the zero value immediately (Figure \ref{fig3} Panel (a)), MSBCs have sharp decreases in correspondence of Rituximab intakes  (Figure \ref{fig3} Panel (b)) and Tregs density is higher as long as low dose IL2 interleukin is administrated  (Figure \ref{fig3} Panel (a)). 
	Thus, we may conclude that the long-term effects of therapy are appreciable only if the minimum value of MSBCs is sufficiently low and the proliferation rate of NBCs is not too small.
	
	\section{Conclusions}\label{seconcl}
	
	A kinetic framework under the action of an external force field has been proposed in this paper and applied to the medical treatment of autoimmune response. 
	
	The kinetic framework has been derived in Section \ref{secmodel}, where a stochastically interacting system is considered, whose microscopic state is described by a continuous activity variable. In particular, the external force field $\mathbf{F}[\mathbf{f}](t,u)$ has a specific shape \eqref{forshap}, in dependence on the application of this paper. In particular, this choice ensures that $\mathbf{F}[\mathbf{f}](t,u)$ depends also on the current state of the whole system, through distribution functions of all functional subsystems. Therefore, the kinetic equations, i.e. a system of nonlinear integro-differential equations, are obtained. Theorem \ref{th1} ensures, at least locally in time, existence, uniqueness, positivity, and boundedness of solution of the related Cauchy problem. These results strictly depend on the shape of the external force field. Nevertheless, this approach may be extended to other choices for $\mathbf{F}[\mathbf{f}](t,u)$.
	
	Section \ref{secappl} provides the application of the kinetic framework \eqref{eqfkin} for the medical treatment of autoimmune response. The results adhere to what is experimentally expected. With respect to previous cases proposed in literature, it is worth pointing out two novelties. The former is the conservative dynamics of NBC activation by interaction with both APCs and MSBCs. The latter is the possibility of modeling a treatment strategy in terms of an external force field. Therefore, this paper may represent a general scheme to model stochastic interacting systems under the action of an external environment. Finally, we have shown that, through a suitable choice of functions and parameters describing the drug delivery, it is possible to predict the diminishing of cell populations responsible for autoimmune dynamics.
	
	It is worth stressing that the specific shape of the external force field is crucial in the current paper. Furthermore, this can be applied also in other contests as, for instance, environmental and socio-economical systems.
	
	However, this work represents a  step at modeling stochastically interacting systems under the action of external force fields. As future research perspectives, results of the global in time existence of the solution have to be considered. In particular, we expect results that depend on the shape of the external force field $\mathbf{F}[\mathbf{f}](t,u)$ and the parameters of the system. Analogously, according to what was recently done in \cite{menale2023kinetic}, nonconservative binary interactions may be considered in order both to have a more realistic depiction of the phenomenon and to gain global in time results. Furthermore, we aim at extending Theorem \ref{th2} in the case of time-dependent parameters of the external force field, i.e. when the system of ordinary differential equations at the macroscopic level is nonautonomous. {The presence of a further nonconservative term could provide specific conditions for deriving a global in time theorem of existence and uniqueness of a positive solution. Roughly speaking, we expect that the action of one nonconservative term could be compensated by the action of the other one.} Finally, from an application view, more realistic scenarios may be considered, for example, a spatio-temporal setting, including the chemotactic motion of cells, as proposed in \cite{immuneProc}, or the introduction of different time scales for the therapy delivery.

	\paragraph*{Acknowledgments}{ The research of the Authors has been carried out under the auspices of GNFM (National Group of Mathematical-Physics) of INdAM (National Institute of Advanced Mathematics). R.T. is holder of a Research Fellowship from the National Institute of Advanced Mathematics (INdAM), Italy. 
    The research work of RT was carried out in the frame of activities sponsored by the Cost Action CA18232 and supported by INdAM (National Institute of Advanced Mathematics), by the Portuguese national funds (OE), through  FCT/MCTES  (Fundação para a Ciência e a Tecnologia) Projects  UIDB/00013/2020,  UIDP/00013/2020, PTDC/03091/2022 (``Mathematical Modelling of Multi-scale Control Systems: applications to human diseases (CoSysM3)”), and by University of Parma through the action Bando di Ateneo 2022 per la ricerca co-funded by MUR-Italian Ministry of Universities and Research - D.M. 737/2021 - PNR - PNRR - NextGenerationEU (project: ``Collective and self-organised dynamics: kinetic and network approaches").}

	\bibliographystyle{unsrt}
	\bibliography{biblioSub}

\end{document}